\documentclass[
reprint,
nofootinbib,
amsmath,amssymb,
aps,
floatfix 
]{revtex4-2}

\usepackage{graphicx}
\usepackage{dcolumn}
\usepackage{bm}
\usepackage[colorinlistoftodos]{todonotes}
\usepackage[export]{adjustbox}%
\usepackage{natbib}
\usepackage{lineno}
\usepackage{amsthm}

\usepackage{xcolor}
\colorlet{rcolor}{black}  

\usepackage{xparse}  

\newif\ifplos
\plosfalse     

\NewDocumentCommand{\maybeplos}{+m +m}{%
	\ifplos #2\else #1\fi
}

\newcommand{\Exp}[1]{\ensuremath{\mathsf{E}\left[#1\right]}}

\renewcommand{\Pr}[1]{\ensuremath{\mathsf{Pr}\!\left\{#1\right\}}}

\DeclareFontFamily{U}{stix2bb}{}
\DeclareFontShape{U}{stix2bb}{m}{n} {<-> stix2-mathbb}{}

\NewDocumentCommand{\indicator}{}{\text{\usefont{U}{stix2bb}{m}{n}1}}

\DeclareMathOperator*{\argmax}{arg\,max}

\theoremstyle{claim}
\newtheorem*{claim}{Claim}

\newcommand{\addedI}[1]{#1}
\newcommand{\deletedI}[1]{}
\newcommand{\replacedI}[2]{#1} 

\newcommand{\addedII}[1]{#1}
\newcommand{\deletedII}[1]{}

\begin{document}
	
	\title{Screening, sorting, and the feedback cycles that imperil peer review}
	
	\author{Carl T. Bergstrom}
	\email{cbergst@u.washington.edu}
	\affiliation{Department of Biology \\ University of Washington \\ Seattle, WA USA} 
	
	\author{Kevin Gross}
	\email{krgross@ncsu.edu}
	\affiliation{Department of Statistics \\ North Carolina State University \\ Raleigh, NC USA \\}
	
	\date{\today}
	
	\begin{abstract}
		\replacedI{Scholarly journals rely on peer review to identify the science most worthy of publication.}{Scholarly publishing relies on peer review to identify the best science.}  Yet finding willing and qualified reviewers to evaluate manuscripts has become an increasingly challenging task, possibly even threatening the long-term viability of peer review as an institution.  What can or should be done to salvage it? Here, we develop mathematical models to reveal the intricate interactions among incentives faced by authors, reviewers, and readers in their endeavors to identify the best science. Two facets are particularly salient. First, peer review partially reveals authors' private sense of their work's quality through their decisions of where to send their manuscripts. Second, journals' reliance on traditionally unpaid and largely unrewarded review labor deprives them of a standard market mechanism---wages---to recruit additional reviewers when review labor is in short supply. We highlight a resulting feedback loop that threatens to overwhelm the peer review system: (1) an increase in submissions overtaxes the pool of suitable peer reviewers; (2) the accuracy of review drops because journals either must either solicit assistance from less qualified reviewers or ask current reviewers to do more; (3) as review accuracy drops, submissions further increase as more authors try their luck at venues that might otherwise be a stretch. We illustrate how this cycle is \replacedI{propelled by}{further propelled by forces including} the increasing emphasis on high-impact publications, the proliferation of journals, and competition among these journals for peer reviews. Finally, we suggest interventions that could slow or even reverse this cycle of peer-review meltdown. 

		\end{abstract}
%
%
	\maketitle
	\maybeplos{\section{Introduction}}{\section*{Introduction}}
When we think about the institution of peer review in science, we often envision reviewers \addedI{as} acting akin to shell collectors, sorting through the fragments of shells on a tropical beach in search of whole specimens worth taking home. Indeed, this is part of what peer review does. But papers don’t simply appear on {\em \maybeplos{Nature}{PLOS Biology}}’s editorial desk the way that shells wash up on an expanse of white sand. Authors choose to put them there---and they make those choices in \replacedI{anticipation}{full awareness} of the evaluation to follow. Given the stringent review process and the wasted time and effort involved in submitting a paper that is eventually rejected, authors screen their own work, targeting appropriate journals rather than sending everything to the most prestigious outlets.  Thus peer review also induces authors to reveal, through their submission decisions, their own private information about the quality of their work \cite{azar2005review, tiokhin2021honest, zollman2024academic}.

However, this service depends on a massive supply of free labor, namely the unpaid and largely uncredited efforts of peer reviewers \cite{bergstrom_free_2001}. But the pool of peer-review labor has become stretched thin, and the problem seems to be worsening. Editors report increasing difficulty in finding reviewers for the manuscripts that they handle. Bibliometric studies in a range of fields support their assertions: reviewers are more likely to decline review invitations and thus the average number of solicitations per acceptance has increased [\cite{albert_is_2016,fox_recruitment_2017,publons_publons_2018,bro2022shared,tropini_time_2023}, though see \cite{zupanc2024becoming} for an exception]. We appear to be in the midst of a ``peer-review meltdown'' in which the peer-review system is becoming woefully overtaxed by the volume of manuscript submissions \cite{alberts_reviewing_2008,arns_open_2014,breuning_reviewer_2015,flaherty2022peer,routledge2025improving,adam_peer_2025}.  


As peer review teeters, scientists have begun experimenting with new ideas \deletedI{and models }to reduce the review load or increase review supply.   \deletedI{Examples abound. } \replacedI{For example, venues}{Platforms} such as Publons and Elsevier's Reviewer Recognition Platform attempt to make reviewing more prestigious by awarding accolades to top reviewers \cite{ravindran2016getting, yu2024can}. Brokerage services tried charging authors to secure peer reviews that could be forwarded to prospective publication \replacedI{outlets}{venues} \cite{stemmle_rubriq_2013,noauthor_axios_nodate}.
These foundered, but a new generation of \replacedI{journal-independent review initiatives}{open-review platforms} such as Review Commons \addedI{and Peer Community In} have emerged in their stead.  Some computer science conferences such as NeurIPS \cite{noauthor_neurips_nodate} keep review loads down by disallowing revision and re-review. 
Elsewhere, some journals\maybeplos{ }{---including {\em PLOS Biology} \addedI{\cite{routledge2025improving}}---}offer portable peer review, where initial reviews follow a manuscript if it is resubmitted to other venues.  Others have suggested tying the opportunity to submit a paper as an author to one's contributions as a reviewer \cite{fox2010pubcreds}, and some journals have \deletedI{even} experimented with cash payments for reviews \cite{chetty2014policies, doble2024bmj, cotton2025effect, gorelick2025fast}.  \addedI{Yet others have studied whether machine review using AI \cite{checco2021ai} and large language models \cite{liang2024can} can complement peer review.  While debate about the propriety of machine review remains unsettled \cite{gruda2025three,bergstrom2025ai}, some reviewers are using LLMs for assistance even when journal or conference guidelines forbid doing so \cite{liang2024monitoring,naddaf2025major,frontiers2025unlocking}.} Some critics have even proposed eliminating prepublication peer-review altogether \cite{heesen2021peer}.  The breadth of these endeavors testifies to scientists' eagerness to place scientific publishing on more stable footing.  

\begin{figure}[!h]
	\begin{center}
		\includegraphics[width = \linewidth]{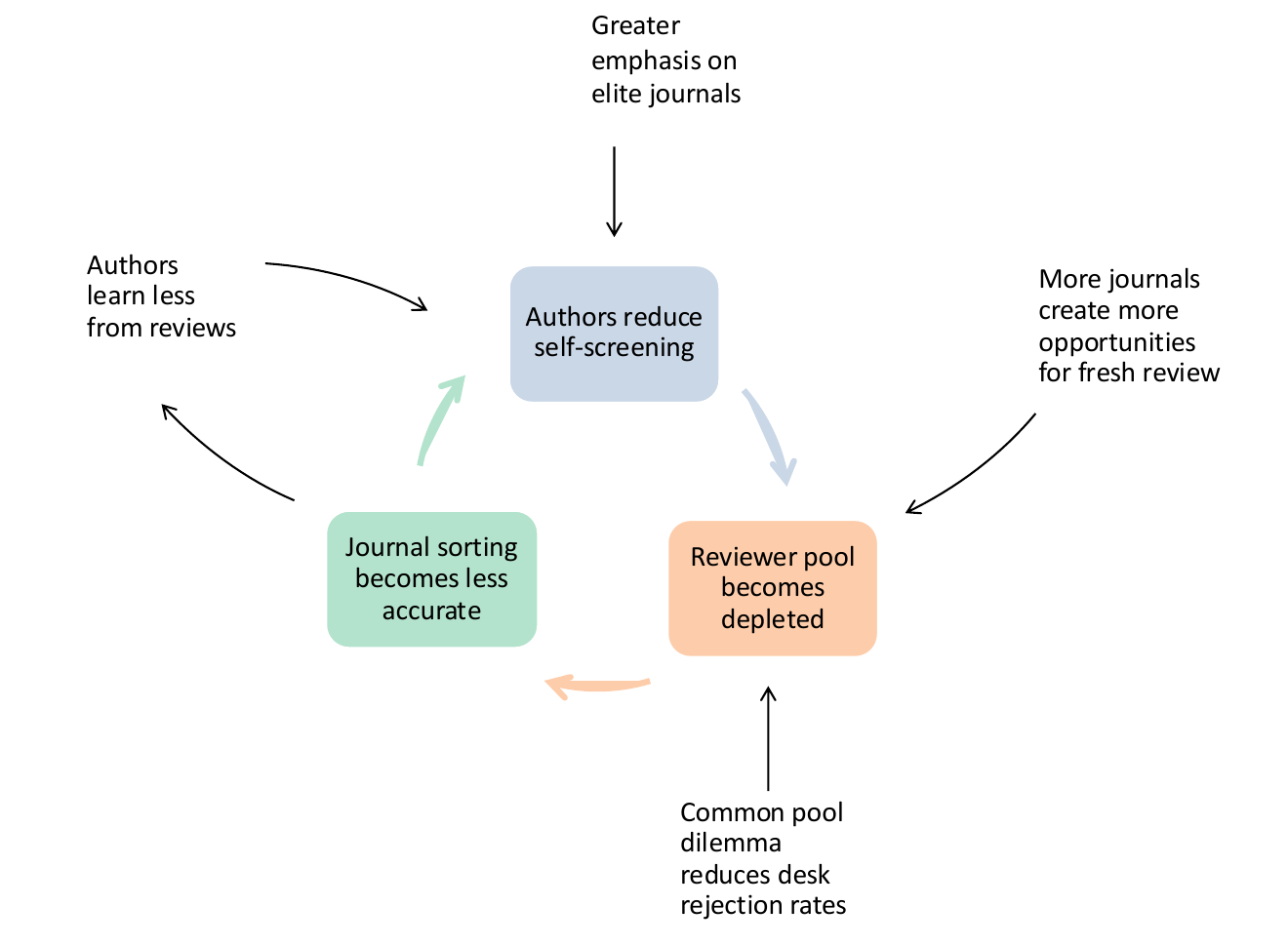}
	\end{center}
	\caption{{\bf The peer-review meltdown cycle.} Author screening and journal sorting interact in a feedback loop which inaccurate \replacedI{sorting}{peer review} loosens author screening \cite{adda2024grantmaking} and looser screening \replacedI{makes sorting less accurate}{leading to less accurate peer review} by depleting the pool of available review labor.   Forces that exacerbate this feedback loop include (clockwise from top): Greater rewards to publishing lead more authors to submit their paper to top journals; a proliferation of journals gives authors more opportunities to obtain fresh reviews of already rejected manuscripts; journals' reliance on a shared pool of review labor compels journals to underuse desk rejection and overexploit the review pool; and noisier review reduces what authors can learn from having their papers rejected\replacedI{.}{, which decreases their selectivity in subsequent submissions.}}
	\label{fig:meltdown}
\end{figure}

Yet all these initiatives are hampered by the fact that a rigorous theory of the structure and function of peer review has yet to coalesce \cite{feliciani2019scoping}.  This article aims to begin to fill that gap.  While the literature is dotted with mathematical models of peer review, many of these efforts use detailed\addedI{,} agent-based \addedI{simulation} models that embrace the richness of the scientific ecosystem (e.g., \cite{thurner2011peer, allesina2012modeling, kovanis2016complex, kovanis2017evaluating, dandrea2017can}).  In this paper, we take a different approach by developing low-dimensional models that \replacedI{isolate}{focus on} how peer review shifts the burden of identifying the best science among among authors, reviewers, and readers of the scientific literature.  (Zhang et al.\ \cite{zhang2022system} have recently presented a model of computer-science conferences that also examines how self-screening by authors affects the peer-review burden.)   These models reveal a set of hidden pressures on peer review, and bringing them to light helps us to understand the tensions that threaten the entire institution as science changes. In particular, these models highlight a pernicious feedback loop: increasing submissions to top journals exhausts the pool of suitable peer reviewers, resulting in lower quality peer reviews that encourage yet more authors to take a chance on submitting their paper to a prestigious venue (Fig. \ref{fig:meltdown}). We also show how other systemic factors---increasing emphasis on high-impact publications, the proliferation of journals, and competition among these journals for unpaid peer-review labor---propel this cycle. Finally, we consider possible solutions that could interrupt the meltdown of peer review, or even reverse it. 

\maybeplos{\section{Model and welfare}}{\section*{Model and welfare}}

\maybeplos{\subsection{Adda-Ottaviani model}}{\subsection*{Adda-Ottaviani model}}  

Our model builds from the base model in Adda \& Ottaviani (2024; henceforth AO) \cite{adda2024grantmaking}.  AO use their model to study grant competitions, but their model adapts naturally to scientific publication.  Our analysis differs substantially from AO, as suits the different aims of the papers.  The \maybeplos{appendix}{S1 Appendix} provides mathematical proofs of several of the key claims and some additional results.

Consider a scientific community served by two journals: an elite journal that seeks to publish \deletedI{the} top \deletedI{$k \in (0,1)$} manuscripts and a mega-journal that publishes everything else.  We consider only two journals for the sake of simplicity, although our model applies in any setting with several vertically differentiated journals.  \replacedI{Suppose that the elite journal has the capacity to publish a proportion $k \in (0,1)$ of the manuscripts that the community produces.}{Call $k$ the elite journal's capacity.}  We assume that \addedI{the journal's capacity} $k$ is determined exogeneously, perhaps by limits imposed by the publisher or by constraints on readers' attention \cite{frankel2022findings}.  Henceforth, we focus on the behavior of the elite journal and place the mega-journal in the background.  Thus we refer to the elite journal as simply ``the journal''; we say that authors who publish their manuscripts in the elite journal are ``published'', and so on.  

Suppose that this community contains a unit mass of authors and that each author is endowed with a manuscript with quality $\theta$.  For mathematical convenience, assume that $\theta$ has a standard Gaussian distribution across authors, $\theta \sim \mathsf{N}(0,1)$.  Authors have their own sense of whether their work is any good. \replacedI{We instantiate this by assuming that an author with a manuscript of quality $\theta$ obtains a private sense of their manuscript's quality $X$ that is drawn from a normal distribution with mean $\theta$ and variance $\sigma_X^2$.}{We instantiate this in the model by assuming that each author receives a private signal of their manuscript's quality.  For an author with a manuscript of quality $\theta$, their private signal $X$ is normally distributed with mean $\theta$ and variance $\sigma_X^2$.} Across authors, \replacedI{$X$ is}{private signals are} marginally distributed as $\mathsf{N}(0, 1 + \sigma_X^2)$.  We refer to the quantile $q=F_X(x)$ as the author's type (here and throughout, $F$ denotes a cumulative distribution function, or cdf), and identify authors with their type, e.g.\ ``author $q$".  

Authors can submit their manuscript to the journal, or not.  Authors who submit their manuscript pay a \deletedI{nonrecoverable} disutility cost $c > 0$ \cite{zollman2024academic}, which includes the opportunity cost of foregoing immediate publication in the mega-journal.  Authors whose manuscripts are published receive kudos, prestige, professional rewards, etc.\ with value $v>c$.  More precisely, $v$ gives the additional reward that the author receives from publishing in the elite journal right away relative to the time-discounted reward of publishing in the mega-journal later.  We ignore any other actions that authors may take (e.g., cover letters) that could communicate information about their type to the journal.

Because the journal cannot directly observe manuscript quality $\theta$, it solicits reviews in the usual way.  For now, assume that journals send out every manuscript they receive for review; desk rejection will be considered later.  Let $Y$ be the review score for a submitted manuscript, and assume that for a manuscript of quality $\theta$, $Y$ is drawn from a Gaussian distribution with mean $\theta$ and variance $\sigma^2_Y$.  Higher review scores $Y$ provide evidence of higher article quality $\theta$, and thus \deletedI{a thresholding rule for $Y$ is rational for a journal that seeks to publish the best manuscripts. Consequently,} the journal \addedI{rationally} publishes those papers whose review score exceeds some acceptance threshold $y$.

Two conditions determine the model equilibrium.  First, authors submit their paper if and only if their payoff from doing so is positive.  We call this the \textit{author-rationality} condition.\maybeplos{\footnote{Indifferent authors will comprise a set of measure zero, so their behavior is irrelevant.  To be concrete, we assume that indifferent authors submit their manuscript.}}{\replacedI{ (Authors on the razor's edge of indifference---those for whom the benefit from submitting their manuscript exactly matches the cost---are rare enough that their behavior does not affect the model equilibrium.)}{(To be concrete, we assume that indifferent authors submit their manuscript, although indifferent authors will comprise a set of measure zero.)}} Second, the journal fills its capacity, a condition we call the \textit{capacity-filling} condition.  The capacity-filling condition can be motivated by assuming that the journal editor prefers to publish as many papers as possible without exceeding the journal's capacity \cite{muller2021gatekeeper}. See ref.\ \cite{zhang2022system} for a related model in which the elite journal is not capacity-limited but instead seeks to publish all papers with a quality above a particular threshold.

\begin{figure}[h!]
	\begin{center}
		\includegraphics[width = \linewidth]{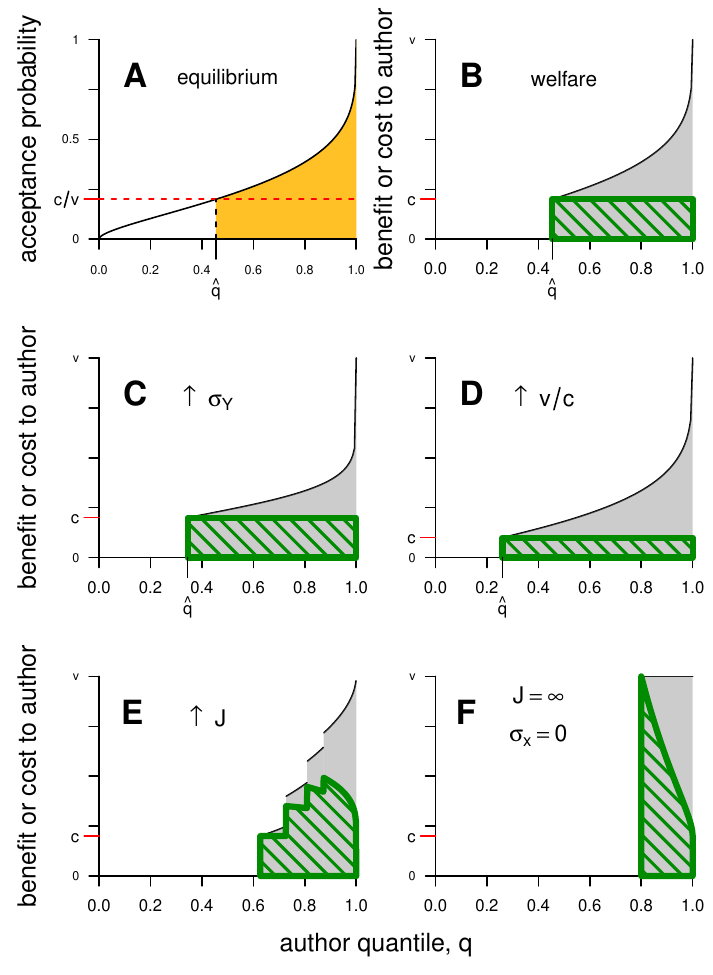}
		\caption{{\bf Graphical illustration of the base model of Adda \& Ottaviani  \cite{adda2024grantmaking} and associated welfare measures.} A: Equilibrium conditions.  The author-rationality condition dictates that the marginal author $\hat{q}$ has acceptance probability $c/v$, and the capacity-filling condition dictates that the area of the gold region---equal to the volume of accepted manuscripts---must equal the journal capacity $k$. Panel based on AO's Figs. 2 and 3 \cite{adda2024grantmaking}.  For this panel, $k=0.2$, $v/c = 5$, and $\sigma_X =\sigma_Y = 1$. B: Welfare measures for the same setting as panel (A). See text for details.  C: Change in welfare under greater peer-review noise (in this case, $\sigma_Y = 2$; all other parameter settings as in panel B). D: Change in welfare as $v/c$ increases (in this case, $v/c = 10$; all other parameter settings as in panel B). E: $J=4$ elite journals.  \addedI{Jumps in the cost and benefit curves show the locations of marginal authors $\hat{q}_i$, $i=1,\ldots,4$.} F: An infinity of elite microjournals and perfect author knowledge of their manuscript's quality.  \maybeplos{}{\addedII{Code to generate this Figure can be found in https://zenodo.org/records/15866736.}}}
		\label{fig:welfare-schematic}
	\end{center}
\end{figure}

Write author $q$'s probability of having their manuscript accepted when facing journal threshold $y$ as $a(q; y) = \Pr{Y \geq y|F_X(X)=q}$.\maybeplos{\footnote{The conditional distribution of $Y$ given $X = x$ is Gaussian with mean $x/(1 + \sigma^2_X)$ and variance $(\sigma^2_X + \sigma^2_Y + \sigma^2_X\sigma^2_Y) / (1 + \sigma^2_X)$.}}{ (The conditional distribution of $Y$ given $X = x$ is Gaussian with mean $x/(1 + \sigma^2_X)$ and variance $(\sigma^2_X + \sigma^2_Y + \sigma^2_X\sigma^2_Y) / (1 + \sigma^2_X)$.)}  Because an author's acceptance probability strictly increases with $q$, there is a marginal author $\tilde{q}$ such only that authors with type $q \geq \tilde{q}$ submit their manuscript.  A model equilibrium is described by a marginal author $\hat{q}$ and a journal cutoff $\hat{y}$ at which the author-rationality and capacity-filling conditions hold.  Writing equilibrium acceptance probabilities as $\hat{a}(q) = a(q; \hat{y})$, the equilibrium solves the author-rationality condition
\begin{equation}
	v \, \hat{a}(\hat{q}) = c
	\tag{AR}
	\label{eq:ar}
\end{equation}
and the capacity-filling condition
\begin{equation}
	\int^{1}_{\hat{q}} \hat{a}(q) \,dq= k.
	\tag{CF}
	\label{eq:cf}
\end{equation}
Note that $v$ and $c$ affect the model only through their ratio.  AO show that the model equilibrium exists, is unique, and is stable.  Fig.~\ref{fig:welfare-schematic}A illustrates the equilibrium's construction.

\maybeplos{\subsection{Welfare}}{\subsection*{Welfare}}

The scientific literature serves (at least) three different constituencies. \textit{Authors} wish to have their work read. \textit{Readers}---and by extension, publishers who market journals to those readers---want to read about the best and most interesting science, and do so in a timely fashion. \textit{Reviewers} contribute the volunteer labor that supports the peer-review process. These constituencies represent roles rather than distinct groups; a given researcher may submit a paper on Monday, read several articles on Tuesday, and write a peer review on Wednesday. But by treating each role as a constituency, we obtain a finer-grained view of how peer review trades off benefits across these activities.  We define payoffs for each group as follows.  

\maybeplos{\subsubsection{Authors}}{\subsubsection*{Authors}}

Author $q$'s payoff is $v \,\hat{a}(q) - c$ if $q > \hat{q}$, and 0 otherwise.  To measure the authors' welfare, we first sum payoffs across authors to give $\int_{\hat{q}}^1 \, [v \,\hat{a}(q) - c] \, dq = v\,k - c\,L$, where $L = 1 - \hat{q}$ gives the volume of papers submitted to the journal, or (in the absence of desk-rejection) the review load. This aggregate payoff simply equals the total rewards from publication, $v\,k$, minus the total disutility cost paid by authors, $c\,L$.  To obtain a scale-free measure of welfare, we standardize this aggregate payoff by $v\,k$, the total payoff available to authors when manuscript quality $\theta$ is public knowledge.  This yields
\begin{displaymath}
	\dfrac{v\,k - c\,L}{v\,k} = 1 - \dfrac{c\,L}{v\,k}
\end{displaymath}
as a measure of the authors' welfare.  

\maybeplos{\subsubsection{Readers}}{\subsubsection*{Readers}}

Readers want to read---and journals want to publish---the best manuscripts.  \replacedI{To this extent,}{Thus} readers' welfare and the journal's payoff are the same, and we measure them by the average quality $\theta$ of published manuscripts (which \replacedI{can be written}{writes} as $\Exp{\theta|q \geq \hat{q}, Y\geq \hat{y}}$) standardized by the average quality of the best $k$ manuscripts ($\Exp{\theta|\theta \geq {F}_\theta^{-1}(1-k)}$).  \replacedI{This measure captures how reliably the journal is able to fill its pages with the best science.  As a measure of reader welfare, it is admittedly incomplete because it ignores the speed with which journals disseminate results; all else equal, readers prefer to learn about new discoveries promptly.}{The readers' welfare measures how reliably publication indicates scientific quality. This measure of reader welfare is admittedly incomplete, because it ignores the speed with which journals disseminate results; all else equal, readers prefer to learn about new discoveries in a timely fashion.}  We lack a metric for reader welfare that incorporates timeliness.

\maybeplos{\subsubsection{Reviewers}}{\subsubsection*{Reviewers}}

While reviewing a manuscript brings a reviewer both benefits and costs, we assume here that the costs \replacedI{of}{to} reviewing exceed the benefits. Thus the burden on the review community \replacedI{scales with}{is proportional to} $L$, the review load.

\vspace*{3ex}

\begin{figure}[h!]
	\begin{center}
		\includegraphics[width = 3in]{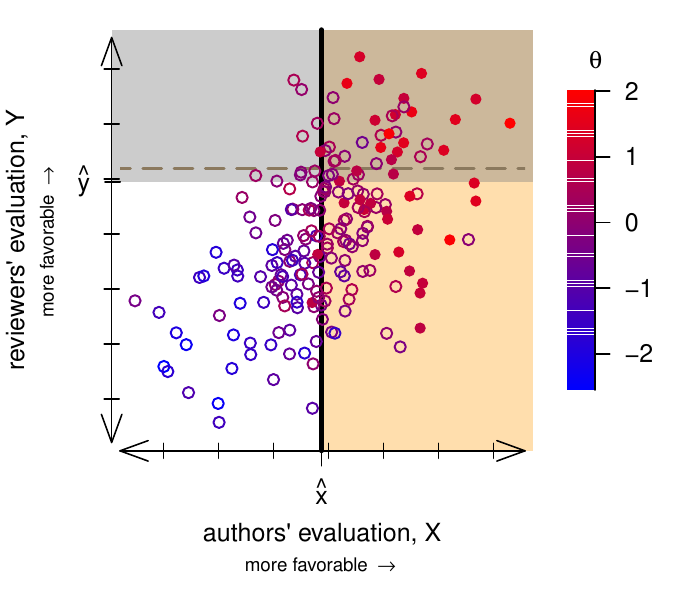}
		\caption{{\bf Author screening and journal sorting combine to \replacedI{determine how capably the journal identifies the most publication-worthy science.}{make the journal and its readers better off than journal sorting alone.}} Points show a random sample of 200 manuscripts, located according to the author's sense of the paper's strength $X$ and by a reviewer's score $Y$.  Point color indicates the quality of the manuscript $\theta$, with redder (bluer) values corresponding to higher- (lower-) quality manuscripts.  The journal publishes $k=20\%$ of the papers, and filled symbols show the actual top 20\% of manuscripts.  At equilibrium, only authors who think their paper is at least as good as $\hat{x}$ (tan region) submit their paper, and the journal then accepts papers that reviewers rate as at least as good as $\hat{y}$ (intersection of tan and gray regions).  The dashed horizontal line shows the acceptance threshold that the journal would use instead if every author submitted their paper. This figure uses $\sigma_x = 0.5$, $\sigma_Y = 1$, and $v/c = 5$. \maybeplos{}{\addedII{Code to generate this Figure can be found in https://zenodo.org/records/15866736.}}}
		\label{fig:trust}
	\end{center}
\end{figure}

Author and reviewer welfare can be understood graphically (Fig.~\ref{fig:welfare-schematic}B). The black curve of Fig.~\ref{fig:welfare-schematic}B gives authors' benefit (for the same parameter values as Fig.~\ref{fig:welfare-schematic}A), $v \,\hat{a}(q)$, and the area under this curve, shaded in gray, equals the authors' total benefit, $v\,k$.  The green hatched region has height $c$ and width $L$, and thus its area equals the authors' total cost, $c \, L$.  Author welfare equals the proportion of the gray shaded area that lies outside the hatched green box.  The load on reviewers equals the area of the green box divided by $c$. \addedI{Reader and journal welfare depends on the actual article quality $\theta$ and thus is determined by the combined action of screening and sorting (Fig.~\ref{fig:trust}).  In \maybeplos{the Appendix (Fig.~\ref{fig:heatmap})}{S1 Figure}, we show that the optimal strength of screening for the journal and its readers depends on authors' and reviewers' accuracy in assessing manuscript quality.  These results are intuitive: journals and their readers are better off with more stringent screening and more selective submissions when authors are well-informed about their paper's quality, and conversely are better off with looser screening and more submissions when referees are better informed.}

\deletedI{Before proceeding, we observe that the journal's publication decisions depend on the combined action of screening and sorting (Fig.~\ref{fig:trust}).  The optimal blend of screening and sorting, at least as far as the journal and its readers are concerned, depends authors' and reviewers' accuracy in judging manuscripts' quality.  For example, if authors know their manuscript quality's perfectly ($\sigma_X = 0$), the journal's payoff rises as screening becomes more stringent and only the top authors submit their manuscript.  Conversely, if reviewers are infallible ($\sigma_Y=0$), the journal prefers that all authors submit their manuscript so that the reviewers can find the best ones.  The more realistic case in which both authors and reviewers are imperfect judges of manuscript quality intergrades between these two extremes, with the journal's optimal blend of screening and sorting set to leverage authors' and reviewers' knowledge in proportion to the accuracy of each (Fig.~\maybeplos{\ref{fig:heatmap}}{A1}).}

\maybeplos{\section{Analysis}}{\section*{Analysis}}

\maybeplos{We begin by analyzing the model with a single elite journal, and then consider a model with several elite journals.  Numerical analysis were conducted using R \cite{R}.\footnote{Code is available at https://doi.org/10.5281/zenodo.15866735.}}{We begin by analyzing the model with a single elite journal, and then consider a model with several elite journals.  Numerical analysis \replacedI{was}{were} conducted using R \cite{R}.}

\maybeplos{\subsection{A single journal}}{\subsection*{A single journal}}

\addedI{A central result of AO is that more authors submit their paper when reviewing is less accurate, and vice versa  (Fig.~\ref{fig:welfare-schematic}C).  The intuition is that less accurate peer review introduces more stochasticity into the journal's sorting. This loosens the relationship between the author's acceptance probability and their type, thus encouraging more authors to submit their manuscripts and relaxing screening.  As a result, less accurate reviewing makes both authors and reviewers worse off by increasing the review load $L$.  The effect on readers' welfare is ambiguous.}

\deletedI{A central result of AO is that more accurate reviewing results in fewer authors submitting their papers, while less accurate reviewing leads to more submissions (Fig.~\ref{fig:welfare-schematic}C).  The intuition is that more accurate peer review tightens the relationship between the author's acceptance probability and their type, which strengthens screening.  Thus more accurate reviewing makes both authors and reviewers better off because it decreases $L$.  The effect on readers' welfare is ambiguous.}

If increasing the review load also decreases the accuracy of peer review, then author screening and journal sorting participate in a feedback loop \cite{zhang2022system}. It is easy to see how an increase in the review load might cause peer review to become less accurate.  Suppose that reviewers differ in their review accuracy and that editors preferentially invite more accurate reviewers.  As the review load increases, editors must either ask their favored reviewers to do more work or they must reach further into the pool of reviewers.  \replacedI{In either case, peer reviews become less accurate, making the journal's sorting less precise.}{In either case, the accuracy of peer reviews declines.}  Thus, review noise $\sigma_Y$ and review load $L$ reinforce each other: more accurate \replacedI{journal sorting}{peer review} compels authors to screen more selectively, and tighter screening results in fewer submitted manuscripts and more accurate peer review. In the other direction, \replacedI{less accurate journal sorting}{noisier peer reviewing} results in looser screening, which in turn creates more work for reviewers, leading to even noisier reviewing.  (A formal mathematical statement of a model with this feedback loop  appears in the \maybeplos{appendix}{S1 Appendix}.)

The feedback between screening and sorting affects how the equilibrium changes when the environment is perturbed.  For example, in recent decades the career rewards from publishing in top-tier journals have increased dramatically \cite{verma_impact_2015,heckman_publishing_2020}.  It is easy to show that when publishing in top journals brings greater rewards---that is, $v/c$ increases---more authors throw their hat in the ring and the review load increases  (Fig.~\ref{fig:welfare-schematic}D). The intuition is obvious: when publication is worth more, authors are willing to take a chance on a lower probability of success.\maybeplos{\footnote{The welfare effects of an increase in $v/c$ are mixed.  Clearly, the increase in the review load imposes an additional burden on reviewers.  The welfare consequences of an increase in $v/c$ to authors are ambiguous, however, because even though more authors are paying the cost of submission, submission is relatively less costly, and thus the net effect could be positive or negative.  The effect of an increase in $v/c$ on the quality of published articles is also ambiguous.}}{}

\begin{figure*}[t]
	\begin{center}
		\includegraphics[width = 6in]{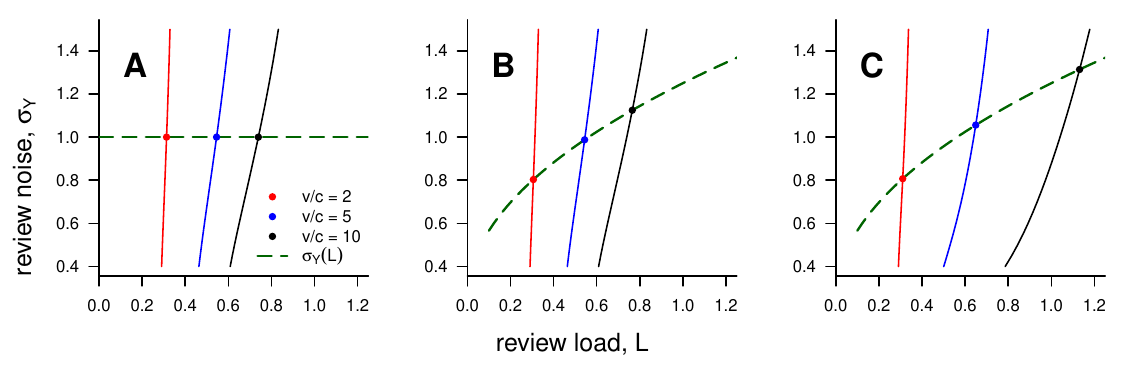}
		\caption{{\bf Feedback effects on peer-review load.} The red, blue, and black curves show the review load $L$ induced by the review noise $\sigma_Y$ for three values of $v/c$.  The dashed green curve shows how the review noise $\sigma_Y$ may respond to the review load $L$.  Equilibria are found at the intersections of the red / blue / black curves with the green curve.  A: A single elite journal and review noise $\sigma_Y$ that is independent of load $L$.  B: A single elite journal and $\sigma_Y$ that increases with $L$.  C: Three elite journals and $\sigma_Y$ that increases with $L$.   Throughout, $\sigma_X = 1$ and $k=0.2$. \maybeplos{}{\addedII{Code to generate this Figure can be found in https://zenodo.org/records/15866736.}}}
		\label{fig:collapse}
	\end{center}
\end{figure*}

The increase in the review load caused by an increase in $v/c$ is exacerbated by the feedback loop between \replacedI{screening and sorting.}{$L$ and the review error $\sigma_Y$.}  Fig.~\ref{fig:collapse}A,B illustrates\replacedI{ by showing how the review load $L$ (determined by the strength of screening) and the review noise $\sigma_Y$ (which determines the accuracy of sorting) shape each other.}{.}  (Fig.~\ref{fig:collapse}C will be considered later.) In each panel of Fig.~\ref{fig:collapse}, the red, blue, and black curves show how the review load $L$ responds to a change in the review noise $\sigma_Y$, for different values of $v/c$.  The dashed green curve shows how $\sigma_Y$ might respond to \deletedI{the review load }$L$.  The equilibrium for a particular \addedI{value of} $v/c$ is found at the intersection of the green curve with the respective red, blue, or black curves.  \deletedI{Comparing Fig.~\ref{fig:collapse}A to Fig.~\ref{fig:collapse}B shows that when} \addedI{When} review accuracy declines with \addedI{increasing} $L$\addedI{ (Fig.~\ref{fig:collapse}B)}, an increase in $v/c$ increases the review load more than it would if reviewer accuracy was independent of $L$\addedI{ (Fig.~\ref{fig:collapse}A)}.  In other words, not only does an increase in $v/c$ compel more authors to submit their paper, the resulting pressure on the review pool decreases the accuracy of review, resulting in yet more authors submitting their manuscripts.\maybeplos{\footnote{In principle, the feedback between sorting and screening could cause the equilibrium to fully unravel, only stopping when every author submits every manuscript.  In Fig.~\ref{fig:collapse}, such an unraveling would occur if the dashed green curve for $\sigma(L)$ intersects the red, blue, or black curves from below.  We don't think such a scenario is likely under reasonable model settings, however, and experience suggests that the peer-review process has not yet completely unraveled.}}{}

The feedback loop between \replacedI{screening and sorting}{$\sigma_Y$ and $L$} also prolongs transient perturbations to the equilibrium caused by one-time shocks.  Although a formal dynamical model is outside the present scope, the intuition for the effect is easily seen.  Suppose there is a one-time shock that \addedI{relaxes screening and} increases the review load to a value $L'$ above its equilibrium level.  If the review noise depends on $L$, then the review noise will increase in response to the surge in submissions to a value $\sigma'_Y$. After the shock is over, the submission load does not immediately return to its pre-shock value; instead, it adjusts to the value $L'' < L'$ induced by review noise $\sigma'_Y$.  The review noise then adjusts to the value $\sigma''_Y$ induced by $L''$, and so on.  Eventually, $L$ and $\sigma_Y$ will return to their pre-shock equilibrium, but these transients will take longer to relax if the review noise temporarily increases in response to the one-time surge in submissions.  This suggests one plausible explanation for why a reported change in reviewer behavior following the COVID-19 pandemic \cite{flaherty2022peer} may be surprisingly persistent.

\maybeplos{\subsection{Several competing journals}}{\subsection*{Several competing journals}}

So far, the model considers a simplified environment with one elite journal and one mega-journal.  Of course, the actual scholarly publishing landscape is much richer.  As science has grown, the number of journals has increased while (mega-journals aside) the proportion of the community's output published by any single journal has declined \cite{bergstrom_economics_2006}.    

Here, we consider how the shift from a few large journals to several smaller ones has impacted scientists' welfare. We focus on ``horizontal'' proliferation of journals within a prestige tier.  ``Vertical'' proliferation, in which journals more densely occupy niches along a prestige spectrum, \replacedI{is more difficult to handle because it introduces greater richness into the author's possible actions and consequently lies}{is} outside the present scope (but see ref.\ \cite{muller2021gatekeeper}). We also abstract away from ways in which journals within a prestige tier may specialize to particular subsets of scholarship.

To study the effects of having several elite journals, we embed our previous model in a richer, multi-journal model.  Instead of a single elite journal with capacity $k$, now suppose there are $J$ elite journals that are identical in all regards and that each have capacity $k/J$. Extend  the model to include an implicit temporal component in which time is divided into discrete periods, and use the previous model to capture authors' and journals' actions within a single period.  In each period, the community (though not necessarily the same authors) generates a unit mass of manuscripts with qualities distributed as $\mathsf{N}(0,1)$.  At each period, authors who have been rejected fewer than $J$ times can resubmit their manuscript to a different journal.  Because the journals are equivalent, authors randomize the order of journals to which they submit their manuscript. Each time an author submits a manuscript, they pay a cost $c$, and they receive a reward $v$ if the manuscript is published. Each time an author is rejected, they rationally become more pessimistic about their manuscript's chances at other, not-yet-tried journals.  To keep the math tractable, we assume that authors only update their beliefs based upon how many times their manuscript has been rejected; they don't observe, or at least don't update their beliefs based upon, the actual review scores $Y$.  Journals do not know the submission history of incoming manuscripts and must treat all manuscripts identically.  

To develop notation, write the conditional probability that author $q$'s manuscript is accepted on the $j$-th attempt when facing review-score threshold $y$ as
\begin{multline*}
	a_j(q;y)
	= \Pr{Y_j \ge y | F_X(X)=q,\;Y_1<y,\dots,Y_{j-1}<y}\\
	j=1,\dots,J\,.
\end{multline*}
While $a_j(q;y)$ does not depend directly on $J$, the equilibrium review cut-off $\hat{y}_J$ will depend on $J$, and thus $a_j(q;\hat{y}_J)$ will as well.  Write the marginal probability that author $q$'s manuscript is accepted on the $j$-th attempt as 
\begin{multline*}
	b_j(q;y)
	= \Pr{Y_1 < y, \ldots, Y_{j-1} < y, Y_j \geq y|F_X(X) = q}\\
	j=1,\dots,J\,.
\end{multline*}
where $b_j(q;y) = 0$ if it is not worthwhile for author $q$ to submit their manuscript a $j$-th time.  Write the probability that author $q$'s manuscript is eventually accepted as $m_J(q;y) = \sum_{j=1}^J b_j(q;y)$.  The total volume of published manuscripts at each period is then $\int_0^1 \, m_J(q;y) \, dq$.  An equilibrium consists of a review-score cutoff $\hat{y}_J$ at which the journals exactly fill their aggregate capacity and a set of marginal authors $\hat{q}_1 \leq \hat{q}_2 \leq \ldots \leq \hat{q}_J$ for which author $\hat{q}_j$ is indifferent about submitting their manuscript for the $j$-th time.  The marginal authors satisfy the author-rationality conditions
\begin{equation}
	v \, a_j(\hat{q}_j, \hat{y}_J) = c 
	\tag{AR-J}
\end{equation}
while the capacity-filling condition \replacedI{is given by}{writes as} 
\begin{equation}
	\int_0^1 \, m_J(q;\hat{y}_J) \, dq = k.
	\tag{CF-J}
\end{equation}

To calculate the review load, write the average number of times that author $q$ submits their manuscript as $\mu_J(q)$, given by
\begin{multline*}
	\mu_J(q)
	= \indicator_{q \ge \hat{q}_1}
	+ \indicator_{q \ge \hat{q}_2} \times (1 - a_1(q,\hat{y}_J)) \\ 
	+ \cdots
	+ \indicator_{q \ge \hat{q}_J} \times \prod_{j=1}^{J-1}\bigl(1 - a_j(q,\hat{y}_J)\bigr).
\end{multline*}
where $\indicator_{q \geq \hat{q}_i}$ is an indicator function that equals 1 if $q \geq \hat{q}_i$ and equals 0 otherwise.  (Note that $\mu_J(q)$ also gives the aggregate density of incoming submissions across all $J$ journals from type-$q$ authors at each period.)  The total volume of submissions, and hence the review load, at each period is then
\begin{displaymath}
	L_J = \int_0^1 \, \mu_J(q) \, dq.
\end{displaymath}

Welfare measures extend naturally.  Reader welfare is given by the average quality of published papers standardized by the quality of the top $k$ papers (a formal expression appears in the \maybeplos{appendix}{S1 Appendix})\replacedI{. The burden on reviewers}{, and reviewer welfare} is proportional to \deletedI{(the opposite of)} the review load $L_J$.  Author $q$'s payoff equals $v \,m(q) -c\,\mu(q)$; integrating over authors again gives $\int_0^1 \, [v \,m(q) -c\,\mu(q)]\, dq = v\,k - c\,L$; thus standardized author welfare is still $1 - c\,L / v\,k$.

\begin{figure*}[t]
	\begin{center}
		\includegraphics[width = 6in]{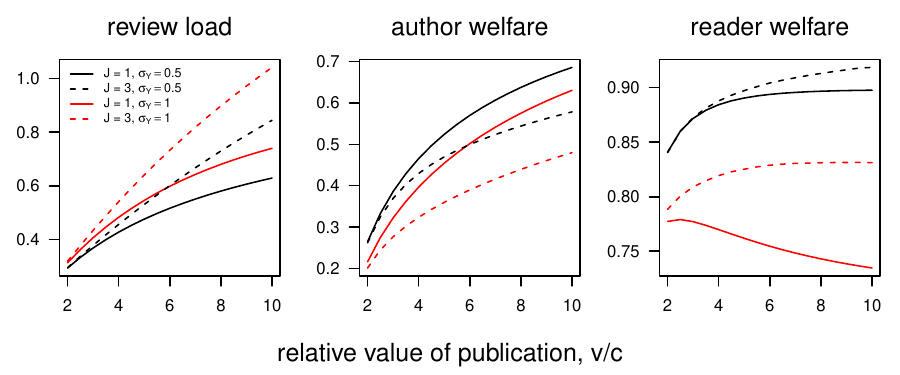}
		\caption{{\bf Welfare consequences of journal proliferation.} Increasing the number of elite journals helps readers but hurts authors and reviewers, and these effects are amplified when peer reviewing is less accurate.  In this figure, the review noise, $\sigma_Y$, is independent of the review load, $L$, to highlight how the number of journals, $J$, interacts with review noise. Left: Review load per period vs. $v/c$ for $J=1$ (solid) or $J=3$ (dashed) journals, with review noise $\sigma_Y = 0.5$ (black) or $\sigma_Y = 1$ (red). Center: Author welfare.  Right: Reader welfare, equal to the (standardized) average value of a published manuscript.  Throughout, $k=0.2$ and $\sigma_X = 1$. \maybeplos{}{\addedII{Code to generate this Figure can be found in https://zenodo.org/records/15866736.}}}
		\label{fig:several-journals-welfare}
	\end{center}
\end{figure*}

The feedback loop between \replacedI{screening and sorting}{the quality of sorting and the strength of screening} carries forward to this multi-journal model, and the underlying logic remains the same.  We focus here on new phenomena caused by the presence of several journals and study \deletedI{the model with }a numerical example. This example suggests that increasing the number of journals results in journals publishing better papers while also increasing the review load and the volume of rejected manuscripts.  Fig.~\ref{fig:welfare-schematic}E shows how this happens.  More journals create more opportunities for authors to have their work considered afresh.  More such opportunities increase the volume of recirculated manuscripts, which in turn forces journals to be more selective. Increased selectivity results in higher rejection rates which increases the volume of recirculated manuscripts, and so on.  Thus, giving authors more bites at the apple strengthens screening in the sense that fewer authors find it worthwhile to submit their paper in the first place ($\hat{q}_1$ increases with $J$), but because those authors can submit their work several times, the sorting burden grows, and the overall volume of submitted (and rejected) manuscripts rises (Fig.~\ref{fig:several-journals-welfare}).  

Moreover, increasing the number of journals has a bigger effect on the review load when reviews are noisy (Fig.~\ref{fig:several-journals-welfare}).  The intuition here is that authors learn less about their manuscript's quality from journal decisions when reviews are noisy than they do when reviews are accurate.  Thus, when reviews are noisy, more authors rationally continue to submit the same manuscript despite previous rejections. Combined with the feedback loop between \replacedI{screening and sorting}{sorting and screening discussed earlier}, this identifies yet another feedback cycle by which the burden on peer reviewers grows: as the number of journals proliferates, authors have more opportunities to have their work considered afresh; more such opportunities increase the load on the review community; an increased review load results in less accurate reviews; declining review accuracy diminishes authors' ability to learn, leading more authors to recycle previously rejected manuscripts, and so on. Fig.~\ref{fig:collapse}C illustrates how an increase in the number of journals amplifies the feedback loop between \replacedI{screening and sorting}{sorting and screening}.

To see the effect of journal proliferation starkly, consider the extreme case in which authors know their manuscript's quality exactly ($\sigma_X = 0$; a similar analysis of this extreme case appears in ref.\ \cite{zhang2022system}).  In this case, an author's acceptance probability at any given attempt is independent of the number of times the manuscript has already been rejected ($a_j(q; y) = a(q;y)$ for all $j$). Thus there is a single marginal author, and an author's eventual probability of acceptance \replacedI{is}{writes neatly as} $m_J(q;y) = 1 - (1-a(q;y))^J$. In the limit as $J$ grows large, any paper that worth submitting gets published eventually ($\lim_{J \rightarrow \infty} m_J(q;y) = 1$).  Writing the limiting marginal author as $\hat{q}_\infty$, the capacity-filling condition becomes
\begin{equation*}
	k = \int_{\hat{q}_\infty}^1 \,  dq  = 1 - \hat{q}_\infty.
\end{equation*}
Thus $\hat{q}_\infty = 1-k$.  

In other words, when perfectly informed authors have an unlimited number of attempts to submit their work, screening is perfect: only the top $k$ authors ever submit their manuscript, and they keep submitting until their manuscript is published (Fig.~\ref{fig:welfare-schematic}F).\maybeplos{\footnote{This might also explain why elite journals are somewhat willing to give top authors an opportunity to appeal for fresh reviews of rejected manuscripts.  At least we've heard that they are.}}{\deletedI{ (This might also explain why elite journals are occasionally willing to give top authors an opportunity to appeal for fresh reviews of rejected manuscripts.)}}  The \textit{entire} function of peer review in this case is to drive \replacedI{the marginal author's}{author $1-k$'s} probability of acceptance down to $c/v$ so that no other authors find it worthwhile to submit.  The sorting function of peer review is just necessary waste, because every submitted manuscript is published eventually.  Further, the equilibrium rejection rate increases with both $v/c$ and $\sigma_Y$.  If $v/c$ rises, the acceptance threshold $\hat{y}_\infty$ must be pushed even higher to discourage non-submitting authors, leading to more superfluous rejection and resubmission of top papers.  If $\sigma_Y$ increases, the acceptance probabilities of all authors above the marginal author decrease, forcing them to submit their paper more often before it is finally accepted.

\maybeplos{\subsection{Desk rejection}}{\subsection*{Desk rejection}}

As screening and sorting weaken, eventually a journal is bound to receive more manuscripts than it can \deletedI{suitably }review.  Of course, journals are not obligated to send every manuscript out for review. Instead, journals can and do counter a surge in submissions by desk-rejecting some manuscripts, thus preserving their available review labor for the most promising submissions \cite{neff2006peer, hadavand_publishing_2024}. Yet anticipating the net effect of desk-rejection is complicated, because, as we have observed, authors choose whether or not to submit their manuscript in anticipation of the sorting process to follow.  Presumably, even an informed journal editor will make less accurate decisions without peer reviews than with them, and so it is unclear on its face how desk rejection will affect screening, and how this will in turn affect the quality of a journal's published articles.  Here, we add desk-rejection to the one-journal model to formalize this idea and briefly explore its potency.  A formal model of desk-rejection with competing peer journals is beyond the scope of the present article, but we informally discuss how competition impinges on desk rejection  later.

\maybeplos{\subsubsection{A single journal}}{\subsubsection*{A single journal}}

To add desk-review to the model with a single elite journal, suppose that the journal editor observes a noisy signal $D$ of the manuscript's quality, with $D \sim \mathsf{N}(\theta,\sigma^2_D)$.  Journals rationally use a threshold rule to decide which papers to reject immediately and which to send out for peer review, with papers sent out for peer-review if $D \geq d$ for some threshold $d$.  To keep matters simple, we assume that manuscripts sent out for review are accepted or rejected based on their review score $Y$ alone. Write author $q$'s acceptance probability when facing desk-rejection threshold $d$ and review threshold $y$ as $a(q; d, y) = \Pr{D \geq d, Y \geq y|F_X(X)=q}$.

For any possible desk-rejection threshold $d$, there is a corresponding marginal author $\hat{q}(d)$ and review score threshold $\hat{y}(d)$ that satisfy the author-rationality and capacity-filling conditions.  The journal's utility $u(d)$ is the average quality of accepted manuscripts, which \replacedI{is given by the expression}{writes as}
\begin{equation*}
	u(d) = \frac{1}{k}\int_{\hat{q}(d)}^1 \, \Exp{\theta|F_X(X) = q, D \geq d, Y\geq \hat{y}(d)} \, a(q; d, \hat{y}(d)) \, dq.
\end{equation*}
At equilibrium the journal sets the desk-rejection cut-off $\hat{d}$ to maximize its utility:
\begin{equation}
	\hat{d} \in \argmax_{d} u(d).
	\tag{JR}
	\label{eq:jr}
\end{equation}
We call this the \textit{journal-rationality} condition.  An equilibrium is given by a triple $(\hat{d}, \hat{q}, \hat{y})$ that solves the \ref{eq:ar}, \ref{eq:cf}, and \ref{eq:jr} conditions.

With desk rejection in play, use $S = 1 - \hat{q}$ to denote the volume of submitted manuscripts, and continue to use $L$ to denote the volume of manuscripts sent out for review.  Author welfare is now given by $1 - c\,S / v\,k$, while \replacedI{the burden on reviewers continues to scale with $L$}{reviewer welfare remains proportional to (the opposite of) $L$}.

\begin{figure*}[t]
	\begin{center}
		\includegraphics[width = 6in]{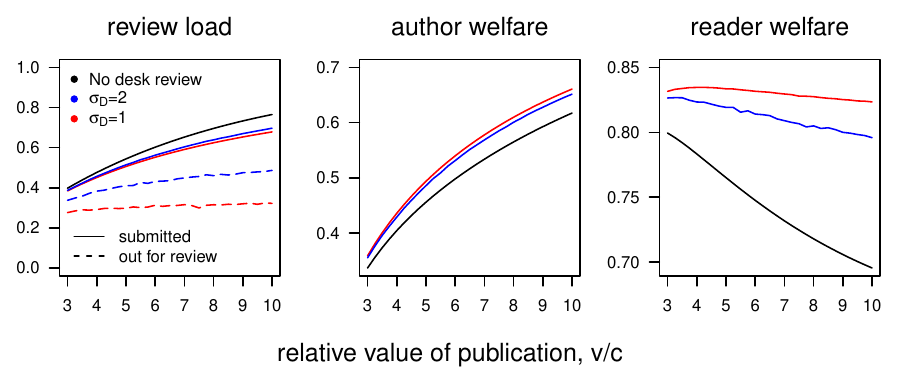}
		\caption{{\bf Welfare consequences of desk rejection.} For a single journal operating in isolation, desk rejection makes all parties---reviewers, authors, and readers---better off.  Welfare measures are shown when desk rejection is absent (black lines), noisy (blue, $\sigma_D=2$), or more precise (red, $\sigma_D=1$).  Left: volume of manuscripts either submitted (solid lines) or sent our for review (dashed lines; submission volume and review load are the same without desk review).  Center: Author welfare.  Right: The readers' welfare, given by the (normalized) average quality of published manuscripts.  For these results, $k=0.2$, $\sigma_x = 1$, and $\sigma_Y = 0.25 + \sqrt{L}$. \maybeplos{}{\addedII{Code to generate this Figure can be found in https://zenodo.org/records/15866736.}}}
		\label{fig:desk-rejection}
	\end{center}
\end{figure*}

Fig.~\ref{fig:desk-rejection} shows how desk-rejection affects the welfare of authors, readers, and reviewers of a single elite journal.  These results show that, at least under the settings explored here, judicious desk-rejection benefits all parties.  Readers (and the journal) benefit because the journal ultimately publishes better papers.  Authors have fewer manuscript rejected and thus recoup more of the available surplus.  Reviewers are faced with fewer manuscripts to review.  Of course, the extent of these benefits depends on the accuracy of the editors' decisions, as captured by $\sigma_D$.  As the accuracy of desk-review declines, its usefulness diminishes accordingly.

\maybeplos{\subsubsection{Several competing journals}}{\subsubsection*{Several competing journals}}

A formal model that includes desk rejection with several competing journals lies outside the present scope, since it would require modeling how authors behave when they face journals with different desk-rejection cutoffs.  However, intuition suggests that when competing journals rely on a common pool of reviewers, a ``tragedy of the reviewer commons'' \cite{hochberg2009tragedy} makes desk rejection less useful for managing the review load.  The logic goes as follows. When a journal \replacedI{has}{operates with} its own private community of reviewers, it fully internalizes the cost of sending a manuscript our for review.  At equilibrium, such a journal sets its desk-rejection policy to optimally trade off the cost of depleting its available review labor against the benefit of using peer review to learn more about a manuscript's quality.  But when several competing journals rely on the same common pool of reviewers, then a journal \deletedI{seeking a review }does not fully internalize the cost of requesting \replacedI{a}{that} review, because the cost of depleting the available review labor is borne by all journals.  Yet the requesting journal still captures the full benefit of obtaining a review and learning more about the manuscript under consideration. Thus, because journals that rely on a shared reviewer pool do not fully internalize the cost of requesting reviews, they will set a desk-rejection policy that desk-rejects manuscripts less often than would be socially optimal. 

In the \maybeplos{appendix}{S1 Appendix}, we formalize this idea with a simple mathematical model whose structure and logic parallel the classic Cournot oligopoly model in economics \cite{cournot1838recherches}.

\maybeplos{\section{Discussion}}{\section*{Discussion}}

Some of the drivers of the peer-review crisis are straightforward and need little elaboration. The number of papers published has been growing about 5\% annually since 1952 \cite{bornmann_growth_2021,hanson_strain_2024}, a rate that exceeds the concomitant growth in the number of university faculty \cite{nces_coe_2024}. Of those faculty, a growing proportion are employed in short-term contingent positions and may be less able to devote time to unpaid and unrecognized review service.  Rapid growth of scientific productivity in non-Western countries has outpaced the fraction of review invitations going to authors located in these regions \cite{publons_publons_2018}, presumably because of mismatches in the composition of editorial boards. These trends have caused peer-review effort to become imbalanced, with a comparably small fraction of researchers providing most of the review labor \cite{kovanis_global_2016,peterson_challenge_2022}.

Meanwhile, both the intrinsic rewards of reviewing and the cost of declining review invitations have shifted.  Pre-print servers and their ilk have eliminated what used to be one of reviewing's major perks, namely the opportunity to obtain an early look at important unpublished work.  Scientific communities are larger and looser knit, making the professional networks that bind editors and reviewers more diffuse.  The universal reliance on e-mail for communication between journals and reviewers has made editors' invitations weaker signals of genuine interest in a reviewer's opinion compared to the manila envelopes and hand-written notes of yesteryear, while evolving social norms reduce the psychic cost to reviewers of ignoring e-mailed invitations altogether. 

Other forces driving the peer-review crisis are subtle and more complex.  We have argued here that the peer-review crisis is partially driven by a pernicious feedback loop in which a growing burden on the reviewer community leads to less accurate reviewing, which in turn compels authors to take more chances with regard to where they submit their work, which exacerbates the burden on the reviewer community further, and so on.  Moreover, the decentralized nature of scientific publishing and the need for journals to compete for authors, readers, and reviewers short-circuits many of the most obvious paths towards halting or reversing this cycle.  While we have presented our model in the context of elite journals, that focus is merely a simplification to streamline the exposition. Similar forces buffet journals across the scientific spectrum.  

Like all models, our model makes simplifications that provide scope for future work.  Perhaps the most substantively, we treat manuscript quality as exogenous.  Of course, authors are not endowed with fully formed manuscripts, but instead they decide how much effort to invest to developing manuscripts based on their anticipation of the journal's scrutiny.  For example, the rise of revise-and-resubmit as an author's likely best outcome may \replacedI{perversely encourage}{have the perverse effect of encouraging} authors to submit manuscripts that are less \addedI{than} polished, both because revision seems inevitable and because journals can no longer expect polished submissions if unpolished submissions become the norm \cite{mcafee2010edifying}.  A model of journal and author behavior that endogenizes manuscript quality would provide more satisfying insight on this count.

If contemporary science has reached a point in which the demand for review labor outstrips the available supply, what (besides increasing desk rejection) might journals do to reconcile the two?  Journals might consider tightening screening by increasing the cost of preparing a manuscript, $c$, thus discouraging more authors from submitting their manuscripts in the first place \cite{tiokhin2021honest}. But monetary fees---whether charged upon submission or acceptance---are more likely to screen for authors with the institutional resources to pay \cite{muller2021gatekeeper}, and costs generated by onerous submission requirements postpone the dissemination of results, thus making journals less marketable.  

Moreover, competition for authors forces journals to keep their costs to authors low.  Competing journals must simultaneously screen authors and compete for them, creating a complex set of incentives. While a proper treatment of the interaction between competition and screening behavior lies outside the present scope (but see ref.\ \cite{muller2021gatekeeper}), intuition suggests that competition for authors interferes with a journal's ability to screen.  For example, if a journal increases its costs $c$, authors will face a trade off between the additional cost of submitting to the now costlier journal vs.\ the better odds of acceptance at that journal resulting from thinner competition for publication slots.  We conjecture that this trade-off will drive top authors who are insulated from the cost of thicker competition away from the costlier journal while attracting bottom authors to it. This adverse selection impels journals to keep their cost of submission low, as we observe throughout the current publishing environment.

\deletedI{Adverse selection for weaker manuscripts also explains why top journals are unlikely to simply tolerate a decline in review quality.  Top authors benefit from accurate reviewing and thus prefer to submit their work to journals that offer it.  Journals who seek to attract top authors must try to keep their reviewing accurate, forcing them to lean heavily on the review pool as the review burden increases.  Again, this seems to describe the current publishing environment, where top journals typically seek more reviews than less prestigious ones.}

If \replacedI{raising}{increasing} submission costs \replacedI{is}{and tolerating less accurate reviews are} off the table, journals might instead try to increase the supply of high-quality review labor.  But journals that seek to recruit more high-quality reviewing are handcuffed by the tradition of peer review as a\deletedI{n} volunteer activity. In a typical labor market, employers can attract more labor by increasing wages. Journals operate without this basic market mechanism when reviewers are unpaid.  If publishers were able to recruit more high-quality review labor by paying appropriate wages, they could maintain constant review quality in the face of an increase in submissions, breaking the feedback loop between screening and sorting.  

To be sure, paying reviewers brings risks.  Non-profit journals would need to pass the cost of reviewer pay on to authors or subscribers, although perhaps society journals could maintain a base of volunteer reviewers by cultivating a sense of shared responsibility for the journal.  Paying reviewers also commodifies review labor, likely irreversibly \cite{gneezy2000fine}.  This commodification could perversely decrease the availability of high-quality review labor if reviewers operating under the honor system perceive their effort to be worth more than the wages that the journal offers \cite{gneezy2000fine, gorelick2025fast, frey2001motivation}.  Nevertheless, recent experiments with modest payments to reviewers have succeeded in accelerating the pace of peer review, at least in the short run \cite{chetty2014policies, cotton2025effect, gorelick2025fast}.  \addedI{Wages needn't be simple direct payments, either.  An intriguing alternative idea is to offer monetary prizes for top reviews, which could sweeten the pot for reviewers while motivating reviewer effort \cite{holmstrom1979moral}.} While substantial trial and error should be expected in establishing a \replacedI{plan}{wage scale} that adequately compensates reviewers, once such a \replacedI{plan}{wage scale} is established, it could give publishers a lever to recruit reviewers that they currently lack.

Are there other options that might salvage the institution of (voluntary) peer review?  We highlight \replacedI{a few}{two} options here, \replacedI{all}{both} of which echo the suggestions of others. First, journals can reduce re-review of revised manuscripts \cite{ellison2002evolving, hadavand_publishing_2024}.  Re-review is currently commonplace: accepted papers frequently go through at least two \cite{huisman_duration_2017} or more \cite{hadavand_publishing_2024} rounds of review prior to acceptance.  But routinely sending revised manuscripts back out for review only burdens the reviewer community further while encouraging both authors and reviewers to modulate their effort in anticipation of the revision and re-review process to follow \cite{mcafee2010edifying}.  Journals can reduce re-review by empowering, or even encouraging, academic editors to be more proactive in their decision-making, especially when evaluating revised manuscripts.  A more radical step to reducing re-review is to take revise-and-resubmit decisions off the table altogether \cite{mcafee2010edifying}.  Second, the scientific community can continue to explore ways of sharing reviews of rejected manuscripts, so that the associated review effort is not wasted.  As our model with several journals suggests, much of the burden on the peer-review community flows from \replacedI{serial re-review}{the need to solicit fresh reviews} of already rejected manuscripts when those manuscripts are sent to new journals.  Cascading review and open-review platforms represent promising moves in this direction.  


\addedI{Finally, one last lever that our model suggests for salvaging peer review is to decrease $v$, the reward that authors reap for publishing in selective journals.  In this model, we have taken the view that $v$ is set exogeneously by the scientific community and, unlike the other solutions we have considered, cannot be changed unilaterally by any single actor.  Further, the forces that determine $v$ are varied and complex because the rewards attached to publication serve a variety of economic and sociological functions in science.  For example, we \cite{gross2024rationalizing} and others \cite{dasgupta1994toward} have argued that rewarding publication constitutes a rational if imperfect scheme for motivating researchers to work hard and to take scientific risks while also protecting researchers' livelihoods from the vicissitudes of scientific chance.  But this work does not consider the pressure that publication rewards place on peer review.  Perhaps a more holistic view would suggest that the scientific endeavor would be better off if fewer rewards flowed to authors who  publish in top outlets.  Or perhaps it wouldn't; the answer remains far from clear.  Regardless, whether science would be better off if the rewards to publication were reduced, and if so, how the scientific community might coordinate to change the prestige associated with publication seem to be rich and increasingly urgent topics to contemplate.}

	\section*{Acknowledgments}
	
	Having written a paper about the demise of peer review, we were heartened to receive three thoughtful, detailed, and on-point reviews of this manuscript. We thank the reviewers for their careful attention and constructive feedback, and for demonstrating that all is not yet lost. We also thank Kyle Myers and Nihar Shah for bringing AO's and Zhang et al.'s papers to our attention.  This work was partially supported by NSF awards SES-2346645 to CTB and SES-2346644 to KG, and by TWCF Diverse Intelligences frameworks grant 32581 to CTB.  KG thanks both the Department of Biology at the University of Washington and the Institute for Advanced Study in Toulouse for visitor support, the latter via funding from the French National Research Agency (ANR) under grant ANR-17-EURE-0010 (Investissements d’Avenir program). \\ \vspace{2mm}
	
	\section*{LLM use statement}
	We used ChatGPT for coding assistance and spot-checking of numerical results, and to find some of the references herein. All of the text and mathematical results are our own human outputs.
	
	\bibliography{meltdown}
	
	\newpage 
	
	\renewcommand\theequation{A.\arabic{equation}}  
	\setcounter{equation}{0}   
	
	\renewcommand{\thefigure}{A.\arabic{figure}}
	\setcounter{figure}{0}
	
	\renewcommand{\thesubsection}{A.\arabic{subsection}}
	\renewcommand{\thesubsubsection}{A.\arabic{subsection}.\arabic{subsubsection}}	
	\onecolumngrid
	\section*{Appendix}
	
	\section*{Appendix: Mathematical proofs and additional results}

\subsection{Proofs}

This section proves some of the main results for AO's model.  The first two proofs below follow similar proofs found in AO \cite{adda2024grantmaking}, and are repeated here for completeness.

\begin{claim}
The equilibrium $(\hat{q}, \hat{y})$ exists and is unique.
\end{claim}

\begin{proof}
Follows AO.  Given a candidate marginal author $\tilde{q}$, author rationality pins down the marginal author's acceptance probability at $c/v$, which in turn pins down the journal's acceptance threshold $y(\tilde{q})$ and thus determines every other author's acceptance probability $\tilde{a}(q) = a(q, y(\tilde{q}))$.  Let $\lambda(\tilde{q}) = \int_{\tilde q}^1 \, \tilde{a}(q) \, dq$ be the volume of accepted manuscripts when the marginal author is $\tilde{q}$.  Clearly, $\lambda(1)= 0$ and $\lambda' < 0$.  Assuming $\lambda(0)\geq k$, there is then a unique solution to the capacity-filling condition $\lambda(q)=k$; this solution is $\hat{q}$ and the associated journal acceptance threshold is $\hat{y}$.  In the unrealistic case that $\lambda(0)<k$, every author submits ($\hat{q} = 0$), the journal decreases $\hat{y}$ to whatever value fills its capacity, and every author has a strictly positive payoff.
\end{proof}

\begin{claim}
The marginal author $\hat{q}$ decreases as review noise $\sigma_Y$ increases.
\end{claim}

\begin{proof}
Follows AO. Let $\hat{q}$ and $\hat{a}(q)$ be the marginal author and acceptance function, respectively, with review noise $\sigma_Y$, and let $\hat{q}_1$ and $\hat{a}_1(q)$ give the marginal author and acceptance function under increased review noise $\sigma_Y' > \sigma$.  Suppose $\hat{q}_1 = \hat{q}$.  Because the marginal author's acceptance probability is pinned down at $c/v$, AO use a result from Lehmann \cite{lehmann1988comparing} to show that $\hat{a}_1(q) < \hat{a}(q)$ for all $q > \hat{q}$.  Graphically, the acceptance probability curve under $\sigma_Y'$ would rotate clockwise around the point $(\hat{q}, c/v)$.  If every investigator $q > \hat{q}$ has a lower probability of acceptance, then the journal wouldn't fill its capacity. Hence thus the journal must lower its acceptance standard, increasing the acceptance probability of the previously marginal author $\hat{q}$ above $c/v$, thus making it worthwhile for more authors to submit papers.
\end{proof}

We make two notes about the proof above.  First, as AO emphasize, the effect of increasing review noise on the journal's cutoff $\hat{y}$ is ambiguous.  However, Lehmann's condition guarantees that, regardless of whether $\hat{y}$ increases or decreases, the acceptance probability $\hat{a}_1(q) < \hat{a}(q)$ for all $q > \hat{q}$, which drives the main result.  Second, the applicability of Lehmann's condition depends on the details of the conditional distribution of $Y$ given $X$ (or alternatively the conditional distribution of $Y$ given $q$).  In our model, Lehmann's condition follows immediately from the fact that $(X,Y)$ has a bivariate normal distribution (which itself follows from our assumption that the triple $(\theta, X, Y)$ has a trivariate normal distribution).  As AO make clear, Lehmann's condition, and hence the above proof, holds more broadly; indeed, AO show that for Lehmann's condition to hold it suffices that the conditional distribution of $Y$ given $X$ has a location-scale structure $Y = X + \sigma_Y \epsilon$ for some random variable $\epsilon$.  However, Lehmann's condition does not necessarily hold universally for all distributions of $(\theta,X,Y)$.  Articulating exactly the collection of distributions on $(\theta,X,Y)$ for which Lehmann's condition holds is beyond the scope of this article.

\begin{claim}
The marginal author $\hat{q}$ decreases as $v/c$ increases.
\end{claim}

\begin{proof}
Fix $c$.  Let $\hat{q}$ and $\hat{y}$ give the equilibrium when the value to publication is $v$, and let $\hat{a}(q)$ give the associated acceptance function. Now consider the equilibrium when the value to publication is $v_1 > v$, and write the associated equilibrium quantities as $\hat{q}_1$, $\hat{y}_1$, and $\hat{a}_1(q)$.  If $\hat{q}$ were to remain the marginal author under $v_1$, then the acceptance threshold would need to be raised to ${\tilde y}_1 > \hat{y}$ to decrease $\hat{q}$'s acceptance probability to $c/v_1$.  But an acceptance threshold of ${\tilde y}_1$ decreases the acceptance probability for every author $q>\hat{q}$, and thus the journal would no longer fill its capacity.  Thus the journal threshold $\hat{y}_1$ must be $<{\tilde y}_1$, so $\hat{a}_1(\hat{q}) > c/v_1$, and hence the new marginal author $\hat{q}_1$ must be $< \hat{q}$. An identical argument holds when $v$ is fixed and $c$ decreases.
\end{proof}

\subsection{Equations for model extensions}

This section gives formal statements of several of the models in the main text. All models build from the AO model \cite{adda2024grantmaking} as described in the main text.

\subsubsection{Peer-review accuracy depends on reviewing load}

Consider a single journal that sends every manuscript that it receives out for review.  Let $L$ denote the review load, such that if the marginal author is $\tilde{q}$, then $L = 1 - \tilde{q}$.  Let the review noise $\sigma_Y$ depend on $L$ through the function $\sigma_Y = \Sigma_Y(L)$.  We assume $\Sigma_Y'(L) \geq 0$.   Write author $q$'s acceptance probability when facing review threshold $y$ with review noise $\sigma_Y$ as $a(q; y, \sigma_Y) = \Pr{Y \geq y|F_X(X)=q}$ where $Y|q \sim \mathsf{N}(F_X^{-1}(q), \sigma^2_Y)$.  Writing equilibrium acceptance probabilities as $\hat{a}(q) = a(q; \hat{y}, \sigma_Y = \Sigma_Y(1 - \hat{q}))$, the model equilibrium can again be found as the solution to AR and CF conditions:
\begin{align*}
v\,\hat{a}(\hat{q}) & = c \\
\int^{1}_{\hat{q}} \hat{a}(q)\,dq & = k. 
\end{align*}
The dependence of $\sigma_Y$ on the review load is included via the new definition of $\hat{a}(q)$.

\subsubsection{Single journal with desk review}

Write author $q$'s acceptance probability when facing desk-rejection threshold $d$ and review threshold $y$ with review noise $\sigma_Y$ as $a(q; d, y, \sigma_Y) = \Pr{D \geq d, Y \geq y|F_X(X)=q}$ where the vector $(D,Y)$ has conditional bivariate normal distribution 
\begin{displaymath}
\begin{pmatrix}
D \\ Y
\end{pmatrix}
\sim \mathsf{N}_2\!\left(
\begin{pmatrix}
x/(1 + \sigma^2_X) \\ x/(1 + \sigma^2_X)
\end{pmatrix},
\dfrac{1}{1 + \sigma^2_X}\begin{pmatrix}
\sigma_X^2 + \sigma_D^2 + \sigma_X^2 \sigma_D^2 & \sigma^2_X \\
\sigma^2_X & \sigma_X^2 + \sigma_Y^2 + \sigma_X^2 \sigma_Y^2
\end{pmatrix}
\right).
\end{displaymath}
Continue to let the review noise $\sigma_Y$ depend on $L$ through the function $\sigma_Y = \Sigma_Y(L)$.  Consider a candidate desk-rejection threshold $d$.  Let $\hat{q}(d)$, $\hat{y}(d)$, and $\hat{\sigma}_Y(d) = \Sigma_Y(\hat{L}(d))$ be the marginal author, review threshold, and review noise induced by $d$, where $\hat{L}(d)$ is the review load induced by $d$ and is given by
\begin{displaymath}
    \hat{L}(d) = \int_{\hat{q}(d)}^1 \, \Pr{D \geq d | F_X(X) = q} \, dq.
\end{displaymath}
The journal's utility $u(d)$ is the average quality of accepted manuscripts, which writes as
\begin{equation*}
    u(d) = \frac{1}{k}\int_{\hat{q}(d)}^1 \, \Exp{\theta|F_X(X) = q, D \geq d, Y\geq \hat{y}} \, a(q; d, \hat{y}(d), \hat{\sigma}_Y(d)) \, dq.
\end{equation*}
This is identical to the expression for $u(d)$ that appears in the main text, with the sole exception that the notation has changed to make the dependence of the acceptance probability on $\hat{\sigma}_Y(d)$ explicit.

\subsection{Additional results}

\subsubsection{Optimal blend of screening and sorting}

\begin{figure}[h!]
	\begin{center}
		\includegraphics[width = 3.5in]{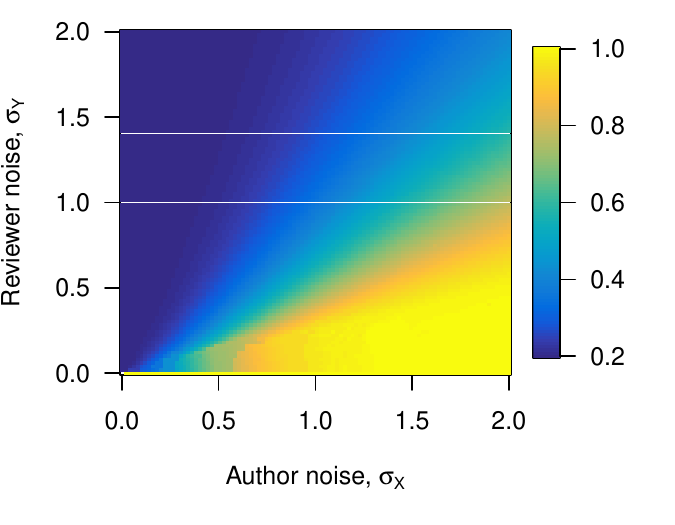}
		\caption{Optimal submission rates. The volume of submissions $L=1-\hat{q}$ that maximizes the quality of published articles as a function of the error in authors' ($\sigma_X$) and reviewers' ($\sigma_Y$) signals of a manuscript's quality. The journal's capacity is $k=0.2$.}
		\label{fig:heatmap}
	\end{center}
\end{figure}

Fig.~\ref{fig:heatmap} shows the submission volume that maximizes the quality of published articles for a journal with capacity $k=0.2$ as a function of the error in authors' private signals and the reviewers signals. The journal wants authors to be more (resp.\ less) selective about submitting their articles when authors' are better (resp.\ worse) judges of their manuscript's quality than reviewers are.  In other words, authors' partial revelation of their private information is more valuable to journals when authors are better than reviewers at assessing their manuscript's true quality, and vice versa.  

\subsubsection{Computing reader welfare}

Computing the reader welfare (or journal payoff) entails computing the mean of a truncated Gaussian distribution.  For example, in the single-journal model without desk rejection, the average quality of a published article is
\begin{equation}
\Exp{\theta | F_X(X)\geq \hat{q}, Y > \hat{y}}.
\label{eq:truncated-gaussian}
\end{equation}
This quantity is then standardized by the average quality of the top $k$ articles  to compute the reader welfare.  Expectations of truncated multivariate Gaussian distributions such as expression \ref{eq:truncated-gaussian} were computed using the \texttt{mtmvnorm} routine from the \texttt{tmvtnorm} package in \texttt{R} \cite{tmvtnorm2023}.

To write the average quality of a published article for the model with several journals, continue with the notation developed there and write the equilibrium volume of manuscripts accepted on their $j$th submission when there are $J$ total journals as 
\begin{displaymath}
\lambda_J(j) = \int_{\hat{q}_j}^1 \, b_j(q, \hat{y}_J) \, dq.
\end{displaymath}
Of course, the capacity-filling condition requires $\sum_{j=1}^J \lambda_J(j) = k$.  Write the average quality of manuscripts accepted on their $j$th submission as 
\begin{displaymath}
\zeta_J(j) = \Exp{\theta|F_X(X)>\hat{q}_j, Y_1 < \hat{y}_J, \ldots, Y_{j-1} < \hat{y}_J, Y_j > \hat{y}_J}.
\end{displaymath}
The average quality of accepted manuscripts is just the weighted average of the $\zeta_J(j)$'s, using $\lambda_J(j)$'s as weights, which writes as
\begin{displaymath}
    k^{-1}\sum_{j=1}^J \zeta_J(j) \lambda_J(j). 
\end{displaymath}

\subsubsection{Underuse of desk rejection from shared reliance on a common reviewer pool}

Here, we present a separate model that demonstrates how competition between journals results in an underuse of desk-rejection and overexploitation of the reviewer pool.  This model and its underlying logic are essentially identical to the classic Cournot oligopoly model in economics \cite{cournot1838recherches}.  This model stands on its own and does not inherit any notation or assumptions from the model presented in the main text.

Ignore screening.  Suppose that, of the manuscripts submitted to a journal, a fraction $\rho \in [0,1]$ are suitable for publication.  Let $\theta \in \left\{0, 1\right\}$ code for a manuscript's suitability, with $\theta=1$ indicating that a manuscript is suitable for publication.  For every suitable manuscript that a journal publishes, it receives a payoff $v_1 > 0$, and for every unsuitable manuscript that it publishes, it receives the negative payoff $v_0 < 0$.  The journal receives a payoff of 0 for rejecting a manuscript.  The fraction of suitable manuscripts is sufficiently low that a journal prefers to reject every manuscript instead of publishing every manuscript, $v_0 (1 - \rho) + v_1 \rho < 0$.

A journal must decide which fraction of its manuscripts to send out for review and which fraction to desk-reject.  Let $L \in [0, 1]$ be the fraction of manuscripts sent out for review, or the review load.  If a manuscript is sent out for review, it is evaluated by a single reviewer who generates a binary report $Y \in \left\{0, 1 \right\}$, where $Y=1$ if the reviewer reports that the manuscript is publishable and $Y=0$ otherwise.  Each reviewer is characterized by a pair of error rates $(\alpha, \beta)$, where $\alpha = \Pr{Y=1 | \theta = 0}$ is the conditional probability that a reviewer incorrectly reports that a manuscript should be published when it is unsuitable, and $\beta = \Pr{Y=0 | \theta = 1}$ is the conditional probability of a report that a suitable manuscript should not be published. 

If a journal sends a manuscript out for review, it must act in accord with the reviewer's report.  If the reviewer reports $Y=1$, the manuscript is published, otherwise the manuscript is rejected.  The value to the journal of sending a manuscript to a reviewer with accuracy $(\alpha, \beta)$ is
\begin{align*}
v(\alpha, \beta) & = v_1 \Pr{\theta = 1, Y = 1} + v_0 \Pr{\theta = 0, Y = 1} \\
& = v_1 (1 - \beta) \rho + v_0 \alpha (1 - \rho). 
\end{align*}
Note that the journal prefers to desk reject a manuscript instead of sending it to a reviewer whose report is independent of the manuscript's suitability (i.e., $v(\alpha, 1 - \alpha)<0$ for any $\alpha$); the journal prefers to send a manuscript to a perfectly accurate reviewer instead of desk rejecting it ($v(0, 0)>0$), and the journal is better off when the reviewer is more accurate (both $\partial v / \partial \alpha <0$ and  $\partial v / \partial \beta <0$).  Because a journal's payoff will be based only on $v(\alpha, \beta)$ and not $\alpha$ and $\beta$ directly, label reviewers by the expected value that they generate for the journal, $v = v(\alpha, \beta)$, i.e., ``reviewer $v$'', and call $v$ the reviewer's value.  

Suppose reviewers differ in their value, and that the distribution of values among reviewers is continuously distributed on an interval contained in $[v_0(1-\rho), v_1 \rho]$. Let $v(q)$ be the value of the reviewer at the $(1-q)$th quantile, such that $v(0)$ is the value of the best reviewer and $v(1)$ is the value of the worst,  with $dv/dq<0$.  A journal's payoff to sending out $L$ manuscripts for review and desk-rejecting the rest is 
\begin{displaymath}
\pi(L) = \int_0^L \, v(q) \, dq = L \bar{v}(L)
\end{displaymath}
where $\bar{v}(L) = L^{-1} \int_0^L \, v(q) \, dq$ is the average value of the top $L$ reviewers.  To maximize its payoff, the journal chooses $L$ to satisfy the first-order condition $\pi'(L) = v(L)=0$.  Let $L^*$ denote this optimal load.  In other words, the journal solicits reviews from all reviewers with $v \geq 0$, and then desk-rejects the remaining manuscripts.  

Now suppose there are $J$ competing journals that all rely on the same common pool of reviewers.  Each journal $j=1, \ldots, J$ receives a randomly selected proportion $1/J$ of the total manuscripts, and must decide what fraction $L_j \in [0, 1/J]$ to send out for review.  If a journal decides to send a manuscript for review, it does not necessarily receive its first choice of reviewers. Instead, if together the $J$ journals send a total of $L$ manuscripts out for review, then each journal obtains its reviews from a random subset of the $L$ best reviewers.  Let $L_j$ be the fraction of its manuscripts that journal $j$ sends out for review, and let $L_{-j}$ give total volume of manuscripts sent out for review by the other $J-1$ journals.  The payoff to journal $j$ is
\begin{displaymath}
\pi_j(L_j, L_{-j}) = \dfrac{L_j}{L}\int_0^L \, v(q) \, dq. 
\end{displaymath}
We seek a symmetric, pure-strategy Nash equilibrium.  Let $L_j^*$ give each and every journal's equilibrium action, and let $L_J^* = J \times L_j^*$ give the total review load at this equilibrium.  Solving the first order conditions $\partial \pi_j / \partial L_j = 0$ and seeking the symmetric solution in which $L^*_{-j} = (J-1) \times L_j^*$ shows that $L_J^*$ solves
\begin{equation}
v(L_J^*) + (J-1)\bar{v}(L_J^*) = 0.
\end{equation}
Because $v(L) < \bar{v}(L)$ for all $L$, it follows that $L_J^*$ is increasing in $J$ (and strictly increasing as long as there is an internal equilibrium.)  

Thus, as $J$ increases, journals desk reject manuscripts more sparingly and send more manuscripts out for review.  More precisely, as $J$ gets very large, the total review load shifts from one in which the marginal reviewer makes the journal no better off than desk rejection ($v(L_1^*) = 0$) and approaches one in which the average reviewer makes the journal no better off than desk rejection ($\lim_{J \rightarrow \infty}\bar{v}(L_J^*) = 0$).

\end{document}